\documentclass[aps,prb,twocolumn,superscriptaddress,
groupedaddress,10pt]{revtex4}
\usepackage{amsmath}
\usepackage{amssymb}
\usepackage{graphicx}
\usepackage{epsfig}
\usepackage{dcolumn}
\usepackage{bm}
\usepackage{bm}
\usepackage{blindtext}
\usepackage{natbib}
\usepackage{siunitx} 

\usepackage[titletoc]{appendix}

\usepackage{simplewick}

\usepackage{bbm}
\makeatletter
\newcommand{\mathleft}{\@fleqntrue\@mathmargin0pt}
\newcommand{\mathcenter}{\@fleqnfalse}
\makeatother

\usepackage{afterpage}

\usepackage[usenames,dvipsnames]{xcolor}
\usepackage{amsthm}
\usepackage{booktabs}
\usepackage{hyperref}
\usepackage{tikz}
\usetikzlibrary{calc}
\usepackage[printwatermark]{xwatermark}
\usepackage{mathtools}

\def\be{\begin{equation}} \def\ee{\end{equation}}
\def\bea{\begin{eqnarray}} \def\eea{\end{eqnarray}}

\def\nn{\nonumber}

\begin{document}
\title{
A shell model for superfluids in rough-walled nanopores
}

\author{Wang Yang}
\affiliation{Department of Physics and Astronomy and Stewart Blusson Quantum Matter Institute, University of British Columbia, Vancouver, B.C., Canada, V6T 1Z1}

\author{Ian Affleck}
\affiliation{Department of Physics and Astronomy and Stewart Blusson Quantum Matter Institute,  University of British Columbia, Vancouver, B.C., Canada, V6T 1Z1}

\begin{abstract}

Recent experiments on the flow of helium-4 fluid through nanopores with tunable pore radius provide a platform for studying the quasi-one-dimensional (quasi-1D) superfluid behaviors.
In the extreme 1D  limit, the helium atoms are localized by disordered small variations in the substrate potential provided by the pore walls.
In the limit of wide pore radius, a solid layer of helium-4 is expected to coat the pore walls smoothing out the substrate potential, and superfluidity is observed in the central region.
Building on earlier quantum Monte Carlo results, we propose a scenario for this crossover using a shell model of coupled Luttinger liquids.
We find that a small radius pore will always localize the helium atoms, but above a critical radius, a single 1D channel flows through the pore and can be described by Luttinger liquid theory.

\end{abstract}
\maketitle

\section{Introduction}
\label{sec:intro}

Superfluidity in bosonic helium-4 can be characterized by flow through narrow pores or constrictions with zero viscosity \cite{Wilks1967,Leggett2006}. 
The walls of such pores are never perfectly smooth, but will always be characterized by some combination of disorder and periodic modulation associated with the solid material through which they traverse. 
Thus, as helium atoms flow through the pore, they will necessarily experience a spatially dependent potential. 
Although the detail of this potential is material dependent, 
its origin lies in the dipole-dipole or Van der Walls interaction between helium atoms and the atoms in the surrounding substrate. 
A ubiquitous feature of such potentials is the presence of a deep potential minima near the surface of the pore. 
This is responsible for the phenomena of wetting \cite{deGennes1985} and drives the escape of superfluid helium from an open container. 
In a confined nanopore geometry, the potential has an approximately cylindrical symmetry,
and the wetting layer will instead form a shell, localized near the pore walls. 
For any excess helium atoms inside this shell, its presence helps to smooth out the localizing effects of disorder or commensuration with the wall,
 and allows for a superfluid component to remain to flow through the center of the pore.
 
On the other hand, a host of recent experiments aim to study the quasi-one-dimensional (quasi-1D) properties of helium-4 confined inside regular nanometer sized constrictions.
Examples of the restricted geometries include
solid helium cells in contact with superfluid helium \cite{Ray2008,Vekhov2012}, 
networks of edge dislocations \cite{Boninsegni2007}, 
and nanopores in mesoporous materials \cite{Sokol1996,Ingaki1996,Dimeo1997,Dimeo1998,Plantevin2001,Plantevin2002,Anderson2002,Toda2007,Taniguchi2011,Savard2011,Prisk2013,Taniguchi2013,Ohba2016,Bryan2017,Bossy2019}.
An alternative approach has been undertaken to study helium-4 mass flow in a single cylindrical nanopore, carved with an electron beam through a thin Si$_3$N$_4$ membrane \cite{Savard2011}.
A major motivation for these experiments is to study the crossover of a quantum fluid to the 1D regime. 
A fluid of interacting bosons at low temperatures ($T$) confined to move along an infinite line is predicted to be a ``Luttinger liquid"  \cite{Haldane1981},
a sort of quasi-superfluid with power law decay at $T = 0$ of the superfluid correlation function 
$\left<\psi^\dagger(x)\psi(y)\right>\propto |x-y|^{-K/2}$ where $\psi(x)$ is the boson annihilation operator. 
Such a liquid is characterized at low energies by its Luttinger parameter $K$, which is a measure of the tendency towards algebraically decaying superfluid or solid order.

\begin{figure}
\includegraphics[width=8.5cm]{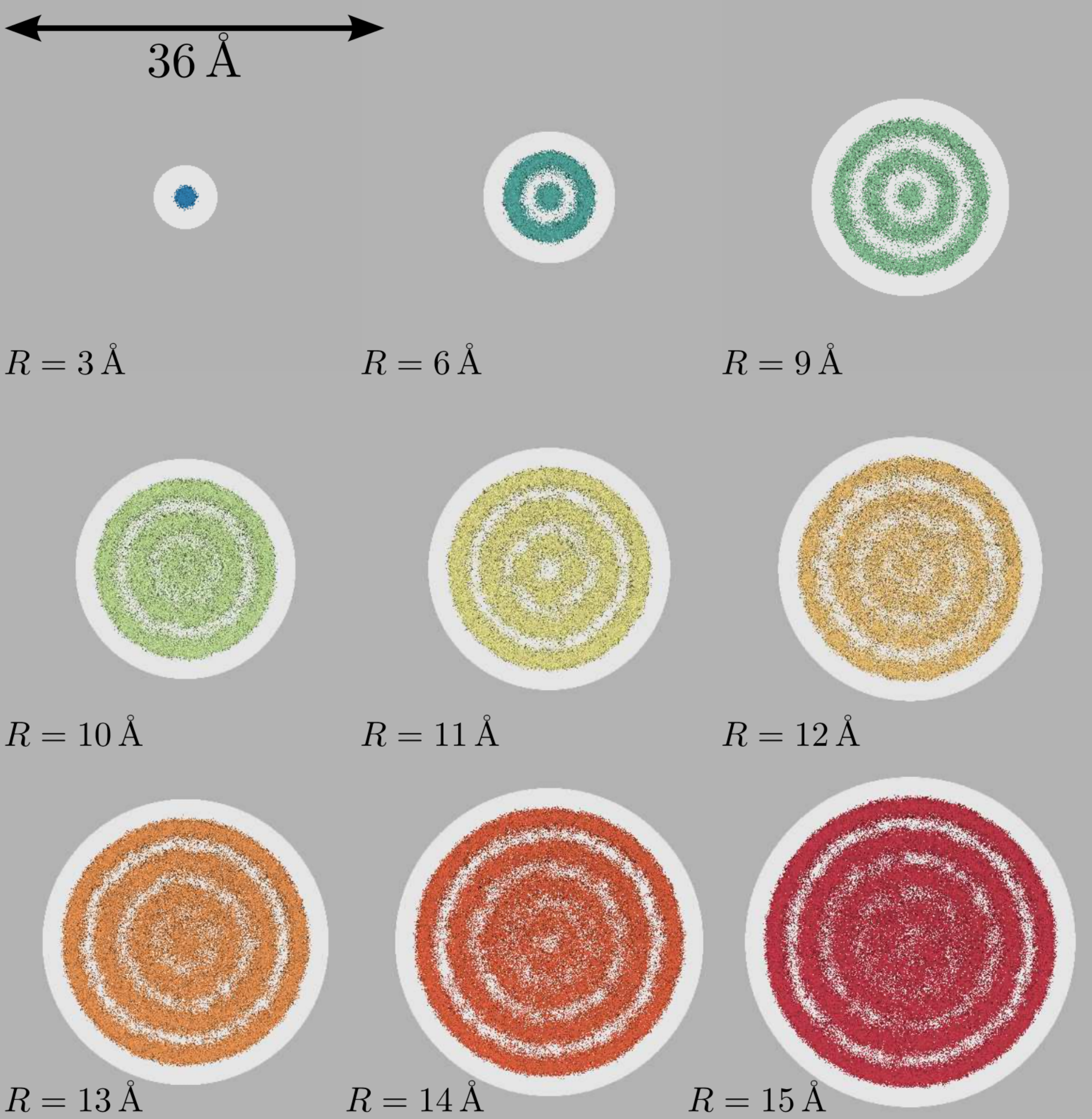}
\caption{Quantum Monte Carlo configurations (particle world lines) projected onto a plane at $T = 0.75 K$ for helium-4 atoms confined inside silicon nitride nanopores with radii between $R = 3\text{--}15\text{\AA}$  and length $L = 10 nm$. 
The azimuthal symmetry in conjunction with a strong confining potential leads to a mass density that oscillates as of a function of radius due to the spontaneous formation of concentric cylindrical shells. 
For full simulation details see Ref. \onlinecite{Maestro2011}.
}
\label{fig:QMC}
\end{figure}

On the theory side, grand canonical quantum Monte Carlo (QMC) simulations  have been performed for helium confined inside smooth nanopores \cite{Maestro2010,Maestro2011},
where realistic interactions between helium atoms and the walls of a translationally invariant Si$_3$N$_4$ pore were included at a chemical potential corresponding to the bulk three-dimensional (3D) saturated vapor pressure.
It was found that a pore of radius $R = 2.9\text{\AA}$          
will support a single quasi-1D column of atoms which can be described at low temperature by Luttinger liquid theory with a large value of $K = 6.0\pm 0.2$ \cite{Maestro2011}.
QMC studies on  smooth cylindrical pores  with larger radii observed the formation of multiple circular layers inside the pore \cite{Maestro2011,Kulchytskyy2013}
and a significantly slower decay of the superfluid correlation function near the pore center.
In Fig. \ref{fig:QMC}, QMC configuration snapshots illustrating this
 behavior are shown for a nanopore with length $L=10nm$ at $T =0.75K$ for $R=3\text{--}15$\AA.

In real Si$_3$N$_4$ pores, 
it is expected that there would be a large confining potential with both periodic and random components due to the glassy structure of the substrate and irregularities in the pore produced by the high energy electron beam.
Even a small external potential is predicted to localize a 1D Luttinger liquid for sufficiently large $K$, with the critical values of $K$ being $1/2$ for a periodic potential commensurate with the helium density and $2/3$ for a random potential \cite{Giamarchi2004}. 
Thus we should expect that experiments on small-radius narrow pores, if possible, would not detect any fluid flow at least for small pressure gradients.
On the other hand, as explained previously,  bulk 3D superfluid behavior is expected when the pore radius reaches the micron range, regardless of the presence of a sizable substrate potential.

\begin{figure}
\includegraphics[width=8.5cm]{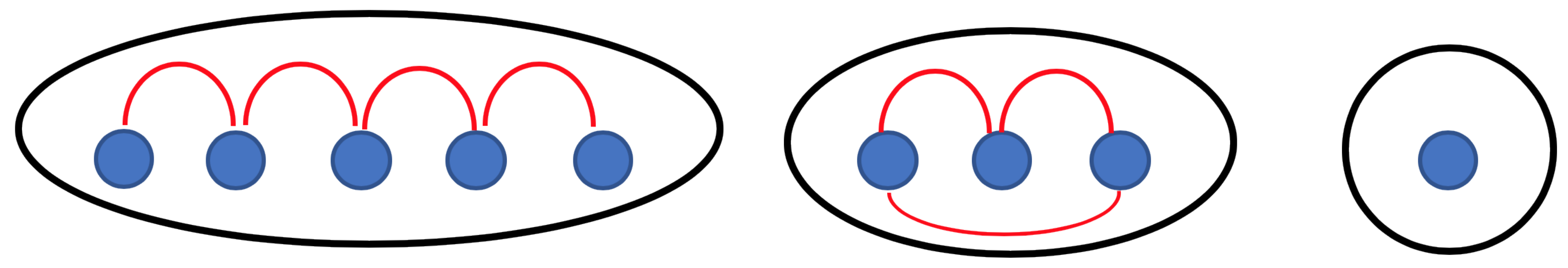}
\caption{Collective groups of channels bound by the relevant inter-shell tunneling operators at low energies.
The solid blue circles represent different Luttinger liquid channels in the absence of inter-shell interactions, inter-shell tunnelings and disorders.
The red line represents the relevant tunneling term between the two channels which the line connects.
The solid black circles represent the group of channels that are bound by the relevant inter-shell tunneling operators in the RG sense.
In this case, $P=9$ and $Q=3$.
}
\label{fig:grouped_channels}
\end{figure}

In this work, based on the results of the QMC simulations in Ref. \onlinecite{Maestro2010,Maestro2011}, 
we develop an analytical theory and propose a shell model of coupled Luttinger liquids to analyze the effects of disordered wall potentials,
 where the Luttinger liquid channels correspond to the shells of concentrated helium atoms as shown in Fig. \ref{fig:QMC}.
Each channel is expected to have a large Luttinger parameter $K$ due to intra-shell interactions. 
The couplings between different channels arise from inter-shell hoppings and the residual inter-shell interactions.
We find that the repulsive  inter-shell interactions always lower the Luttinger parameters, at least for small  interaction strengths.
If the Luttinger parameters are rendered small enough, 
the inter-shell hoppings become relevant at low energies in the sense of renormalization group (RG), and are able to pin the superfluid phases of the corresponding shells among which the hoppings take place.
As a result, the $P$ Luttinger liquid channels are regrouped  into $Q$ ($\leq P$) bound entities,
such that at low energies, the channels within each entity share a same  superfluid phase.
An illustration of such regrouping is shown in Fig. \ref{fig:grouped_channels}.
In particular, the Luttinger parameters of the groups of bound channles will be significantly lowered, making them more immune to disorder effects.

Based on this analysis, we propose a scenario of the crossover behavior from narrow to wide nanopores: the helium atoms are localized by the random substrate potential for small radius pores, whereas there exists a critical radius value, above which a  single (grouped) channel Luttinger liquid emerges in the central region of the nanopore.
This scenario indicates that the large Luttinger parameter in 1D helium-4 does not necessarily destroy the hope of observing the 1D to 3D crossover in experiments and may actually make it easier due to the resulting increase in the critical pore radius.
It would be desirable to compare the predictions here with QMC simulations as well as real experiments. 

Finally, we also note the whole analysis does not necessarily rely on the decomposition of channels based on cylindrical shells. 
Other ways of choosing the channels, for example, angular momentum decomposition, work equally well, which is discussed in Sec. \ref{sec:center_of_mass}.

The rest of the paper is organized as follows.
In Sec. \ref{sec:model_ham}, the model Hamiltonian is introduced and the bosonization is performed. 
In Sec. \ref{sec:intershell_interactions}, the effects of inter-shell interactions on the scaling dimensions of the inter-shell tunnelings are analyzed.
In Sec. \ref{sec:low_energy_field_theory}, the low energy theory for the Hamiltonian including the inter-shell interactions and inter-shell tunnelings is derived.
Based on the results in the previous sections, Sec. \ref{sec:substrate} discusses the effects of disordered substrate potential.
Finally in Sec. \ref{sec:conclusion}, we briefly summarize the main results of the paper.

\section{The model Hamiltonian}
\label{sec:model_ham}

It is a familiar idea that the single particle quantum wave-functions in an infinitely long small radius pore correspond to a set of sub-bands with different transverse wave-vectors. However, that is not the approach we are using here. 
As indicated in Fig. \ref{fig:QMC}, for pore radii of $4\text{\AA}$ or greater, several concentric cylindrical shells of helium atoms form inside the nanopore. 
This is a consequence of the Aziz potential \cite{Aziz1979} describing the interaction between helium atoms and also the potential used to model the interaction with the smooth wall of the pore. 
The density of helium atoms is suppressed at radii between the shells, motivating a starting point in which tunneling (i.e., hopping) of atoms between shells is ignored. 

Then, for suitably long pores, each shell may be considered as an independent 1D system, giving a P-channel Luttinger liquid for a pore with P shells. 
Each of these shells will have a different linear density of atoms and different effective 1D inter-atomic interactions. 
At least three effects need to be included if this model is to be used to describe the physics of quantum fluids in real nanopores: inter-shell interactions, inter-shell tunneling, and the substrate potential which we might expect to be larger on the outer shells near the pore wall than on the inner shells. 
This model corresponds to a multi-leg ladder, with each leg corresponding to a shell.

We consider the Hamiltonian of $P$ channels of Luttinger liquids, as 
\bea
H=H_0+H_{\text{int}}+H_T,
\label{eq:H}
\eea 
in which: $H_0$ is the sum of intra-channel terms
\bea
H_0=&\sum_{i=1}^P\big[ \int dx \psi_i^\dagger (-\frac{\hbar^2}{2m}\frac{d^2}{dx^2}-\mu_i)\psi_i\nn\\
&+\frac{1}{2}\int dxdy\rho_i(x)\hat{U}_{ii}(x-y)\rho_i(y)\big],
\eea
where $\psi_i^\dagger$, $\mu_i$  and $\rho_i(x)=\psi_i^\dagger(x) \psi_i(x)$
are the boson creation operator, the chemical potential and the density operator, respectively, in the $i$'th channel;
$H_{\text{int}}$ includes the inter-shell density-density interactions  as
\bea
H_{\text{int}}=\sum_{1\leq i<j\leq P} \int dx dy \rho_i(x)\hat{U}_{ij}(x-y)\rho_j(y);
\label{eq:hatU}
\eea
and $H_T$ is the inter-shell tunneling term
\bea
H_T=\sum_{1\leq i<j\leq P} \int dx dy \big[t_{ij}(x-y) \psi_i^\dagger (x)\psi_j(y)+\text{h.c}.\big],
\label{eq:tunneling}
\eea
where ``h.c." is ``hermitian conjugate" for short.
Later we will also include the substrate/disorder potential $H_S$ given by
\bea
H_S=\sum_{i=1}^P \int dx V_i(x)\rho_i(x),
\eea
in which $V_i(x)$  represents the substrate/disorder potential acting in the $i$'th channel.

The above Hamiltonians can be expressed in  bosonized forms.
We introduce the bosonization fields $\theta_i(x)$, $\phi_i(x)$, such that $\psi_i(x)$ and $\rho_i(x)$ can be expressed in terms of  $\theta_i(x),\phi_i(x)$ using the following bosonization formulas \cite{Haldane1981},
\begin{flalign}
&\psi_i(x)=\sqrt{\rho_{0i}+\frac{1}{\pi}\frac{d\theta_i(x)}{dx}}e^{-i\phi_i(x)},\nn\\
&\rho_i(x)= \rho_{0i}+\frac{1}{\pi} \frac{d\theta_i(x)}{dx}+[\text{const.} \times e^{2\pi i \rho_{0i}x-2i \theta_i(x) }+\text{h.c.}],\nn\\
\end{flalign}
in which $\theta_i(x),\phi_i(x)$ satisfy the commutation relations
\bea
[\theta_i(x),\phi_j(y)]=\frac{\pi}{2}\delta_{ij}\text{sgn}(x-y),
\label{eq:commutation}
\eea
and  $\rho_{0i}$ is the average density in the $i$'th channel.

After bosonization, $H_0$ acquires the form
\bea
H_0=\frac{1}{2\pi} \int dx \sum_{i=1}^P \big\{ v_{Ji} \big[\frac{d\phi_i(x)}{dx}\big]^2+v_{Ni} \big[\frac{d\theta_i(x)}{dx}\big]^2
\big\},
\eea
in which the Luttinger parameter $K_i$ and the velocity $v_{si}$ are related to $v_{Ni}$ and $v_{Ji}$ by
\bea
K_i= \sqrt{\frac{v_{Ni}}{v_{Ji}}},~v_{si}=\sqrt{v_{Ni}v_{Ji}}.
\eea
For later convenience, we write $H_0$ in a matrix form
\bea
H_0=\frac{1}{2\pi} \int dx \nabla \theta^T V_N \nabla \theta+\frac{1}{2\pi} \nabla \phi^T V_J \nabla \phi,
\label{eq:H0_matrix}
\eea
in which $\theta$ and $\phi$ are both $P$-components column vectors defined as
\bea
\theta=(\theta_1,...\theta_P)^T, ~~\phi=(\phi_1,...\phi_P)^T,
\eea
and $V_J$, $V_N$ are diagonal matrices whose matrix elements are given by
\bea
(V_N)_{ij}=v_{Ni}\delta_{ij},~~(V_J)_{ij}=v_{Ji}\delta_{ij}.
\label{eq:VN_VJ}
\eea

The inter-shell interaction term $H_{\text{int}}$ acquires the bosonized form
\bea
H_{\text{int}}=\frac{1}{\pi^2} \sum_{i<j=1}^P \int dx \hat{U}_{ij}\frac{d\theta_i}{dx}\frac{d\theta_j}{dx},
\eea
in which we have only kept the local terms, and the oscillating terms in the density operators drop off the expression under the assumption that different channels have different densities $\rho_{0i}$.
$H_{\text{int}}$ can also be written in a matrix form
\bea
H_{\text{int}}=\frac{1}{\pi^2}\int dx \nabla \theta^T U \nabla \theta,
\label{eq:Hint_matrix}
\eea
in which the matrix elements of $U$ are given by
\bea
U_{ij}=(1-\delta_{ij}) \hat{U}_{ij}.
\eea
Notice that unlike $\hat{U}$ in Eq. (\ref{eq:hatU}), the diagonal matrix elements of $U$ are all zero.
Here we note that  the two-leg version of this model was studied in Ref. \onlinecite{Orignac1998}
in the special case where the two legs are equivalent, having equal densities and velocity parameters. 
In our case, the occurrence of different densities actually simplifies the analysis since the coupling of oscillating density operators $\cos\{2\pi (\rho_{0i} - \rho_{0j})x - 2[\theta_i(x) - \theta_j(x)]\}$ can be dropped from the low energy theory due to the oscillating phase.

The inter-shell tunneling terms in Eq. (\ref{eq:tunneling}) acquires the bosonized form
\bea
H_T=\sum_{1\leq i<j\leq P}t_{ij}\int dx \cos(\phi_i-\phi_j).
\eea

Finally, the disorder term can be  bosonized as 
\bea
H_S=\sum_{j=1}^P \int dx V_j(x) \cos [2\pi \rho_{0j}x -2\theta_j(x)].
\label{eq:H_S}
\eea

\section{The inter-shell interactions}
\label{sec:intershell_interactions}

In this section, we consider the effects of inter-shell interactions.
As will be discussed in Sec. \ref{subsec:canonical}, the Hamiltonian remains quadratic by including the inter-shell interactions and can be diagonalized by performing a canonical transformation.
Then in Sec. \ref{subsec:scaling_dim}, we determine the scaling dimensions of the inter-shell tunneling terms.

\subsection{Canonical transformation}
\label{subsec:canonical}

Including the inter-shell interactions, the Hamiltonian $H_1=H_0+H_{\text{int}}$ becomes
\bea
H_1=\frac{1}{2\pi}\int dx \big[\nabla \theta^T(V_N+\frac{1}{\pi}U)\nabla \theta +\nabla \phi^T V_J \nabla \phi
\big].
\eea

Define $\theta^\prime$ and $\phi^\prime$ as
\bea
\phi^\prime=V_J^{1/2} \phi,~\theta^\prime=V_J^{-1/2} \theta,
\label{eq:prime_transform}
\eea
then $H_1$ can be written as 
\begin{flalign}
&H_1=\frac{1}{2\pi} \int dx \big[\nabla \theta^{\prime,T}V_N^\prime \nabla \theta^\prime+ \nabla \phi^{\prime,T} \nabla\phi^\prime
\big],
\end{flalign}
in which 
\bea
V_N^\prime=V_J^{1/2} (V_N+\frac{1}{\pi}U) V_J^{1/2},
\eea
and the matrix kernel of the $\phi^\prime$ term becomes the $P\times P$ identity matrix.

Let $O$ be an orthogonal matrix that diagonalizes $V_N^\prime$ i.e.,
\bea
V_N^\prime =O \Lambda_\theta O^T,
\label{eq:diagonalize_VN_U}
\eea
where $\Lambda_\theta$ is a diagonal matrix,
and define 
\bea
\phi^{\prime\prime}=O^T \phi^\prime,~\theta^{\prime\prime}=O^T \theta^\prime,
\label{eq:prime_transform_2}
\eea
then we obtain 
\bea
H_1=\frac{1}{2\pi}\int dx \big[ 
\nabla \theta^{\prime\prime,T} \Lambda_\theta \nabla \theta^{\prime\prime}
+\nabla \phi^{\prime\prime,T}  \nabla \phi^{\prime\prime}
\big].
\eea
The Luttinger parameter $K^{\prime\prime}_j$ in the $j$'th channel  is given by the $j$'th eigenvalue of $V_N^\prime$, i.e.,
\bea
K^{\prime\prime}_j=\sqrt{(\Lambda_\theta)_{jj}}.
\label{eq:Kprimeprime}
\eea
Here we note that the Luttinger parameter $K^{\prime\prime}_j$ is not dimensionless.
This is because after the transformation Eq. (\ref{eq:prime_transform}), 
the new coordinates $\phi^\prime, \theta^\prime$ acquire dimensions, unlike the original canonical coordinates $\phi, \theta$  which are dimensionless.
Alternatively, one can introduce an arbitrary velocity  $v_0$ into Eq. (\ref{eq:prime_transform}), such that the transformations become 
$v_0^{1/2}\phi^\prime=V_J^{1/2} \phi$ and $v_0^{-1/2}\theta^\prime=V_J^{-1/2} \theta$.
Then $K^{\prime\prime}_j$ becomes dimensionless which is dependent on the scale $v_0$. 
However, we will keep using Eq. (\ref{eq:prime_transform}) in this paper for simplification of notations, since this does not affect any physical observable.

\subsection{Scaling dimensions of inter-shell tunnelings}
\label{subsec:scaling_dim}

The scaling dimension of the field $e^{i\lambda \phi_j^{\prime\prime}}$  is \cite{Giamarchi2004}
\bea
[e^{i\lambda \phi_j^{\prime\prime}}]= \frac{1}{4}\lambda^2 K_j^{\prime\prime},
\eea
where $[...]$ denotes the scaling dimension of the operator inside the bracket.
Thus, to get the scaling dimensions of the tunneling terms $\cos(\phi_i-\phi_j)$'s, we need to rewrite $\phi_i-\phi_j$ in terms of $\phi_j^{\prime\prime}$.

Denote $\hat{e}_i$ to be the $P$-dimensional unit column vector along the $i$'th direction, i.e.,
\bea
\hat{e}_i=(0,...0,1,0,...0)^T,
\label{eq:e_i}
\eea
in which $1$ appears at the $i$'th position.
Let 
\bea
x_{ij}=\hat{e}_i-\hat{e}_j,
\eea
then $\phi_i-\phi_j=\phi^T x_{ij}$.
Using Eqs. (\ref{eq:prime_transform},\ref{eq:prime_transform_2}), we obtain
$\phi^T x_{ij}=\phi^{\prime\prime,T} y_{ij}$,
in which 
\bea
y_{ij}=O^T V_J^{-1/2}x_{ij}.
\label{eq:y_x_ij}
\eea

Notice that the scaling dimension of $\cos(\phi^{\prime\prime,T}y)$ is 
\bea
d_{ij}= \frac{1}{4} \sum_{l=1}^P [(y_{ij})_l]^2K_l^{\prime\prime}=\frac{1}{4}y_{ij}^T \Lambda_\theta^{1/2} y_{ij},
\eea
where $(y_{ij})_l$ is the $l$'th component of the column vector $y_{ij}$,
and $K_l^{\prime\prime}$ is given by Eq. (\ref{eq:Kprimeprime}).
Using  Eqs. (\ref{eq:diagonalize_VN_U},\ref{eq:y_x_ij}), we obtain
\bea
d_{ij}(U)=\frac{1}{4} x_{ij}^T V_J^{-1/2} \sqrt{(V_s)^2+\frac{1}{\pi}V_J^{1/2}U V_J^{1/2}} V_J^{-1/2} x_{ij},\nn\\
\eea
in which 
$V_s=\sqrt{V_NV_J}$.

Since $V_s^2$  in general does not commute with $\frac{1}{\pi}V_J^{1/2}U V_J^{1/2}$,
the square root $\sqrt{V_s^2+\frac{1}{\pi}V_J^{1/2}U V_J^{1/2}} $ cannot be easily carried out.
We will consider this square root in the limit of a small $U$, and only keep the results up to first order in the matrix elements $U_{ij}$.
To proceed, the following lemma is needed and a proof is included in Appendix \ref{sec:Proof_lemma}.\\

{\it Lemma.} 
Let $A$ and $B$ both be real symmetric matrices. 
Suppose $A$ is also positive definite.
Then 
\bea
\big[\frac{d}{d\lambda} \sqrt{A+\lambda B} \big]_{\lambda=0}=\int_0^\infty dt e^{-t\sqrt{A}} B e^{-t\sqrt{A}}.
\label{eq:D_integral}
\eea\\

Now we apply Eq. (\ref{eq:D_integral}) to our case.
By taking $A=V_s^2$ and $B=\frac{1}{\pi}V_J^{1/2} UV_J^{1/2}$, we obtain
\begin{flalign}
&\sqrt{V_s^2+\frac{1}{\pi}V_J^{1/2} UV_J^{1/2}}-V_s\nn\\
&=\int_0^\infty dt e^{-tV_s} (\frac{1}{\pi}V_J^{1/2} UV_J^{1/2}) e^{-tV_s}+O(U^2).
\end{flalign}
Thus, to linear order in $U$, $\Delta d_{ij}=d_{ij}(U)-d_{ij}(U=0)$ can be expressed as
\begin{flalign}
&\Delta d_{ij}=\nn\\
&\frac{1}{4}x_{ij}^T V_J^{-1/2}\big[\int_0^\infty dt e^{-tV_s} (\frac{1}{\pi}V_J^{1/2} UV_J^{1/2}) e^{-tV_s}\big] V_J^{-1/2}x_{ij}.
\end{flalign}
Notice that both $V_s$ and $V_J$ are diagonal matrices, hence they commute.
Using the expressions for $x_{ij}$, we obtain
\bea
\Delta d_{ij}&=&\frac{1}{4\pi}\int_0^\infty dt 
x_{ij}^T e^{-tV_s} Ue^{-tV_s}x_{ij}\nn\\
&=&-\frac{U_{ij}}{2\pi}\int_0^\infty dt e^{-t(v_{si}+v_{sj})}\nn\\
&=&-\frac{U_{ij}}{2\pi(v_{si}+v_{sj})},
\eea
in which $v_{sj}=(V_s)_{jj}$.

In summary, the scaling dimension of $\cos(\phi_i-\phi_j)$ is given by
\bea
d_{ij}(U)=\frac{1}{4}(K_i+K_j)-\frac{U_{ij}}{2\pi(v_{si}+v_{sj})} +O(U^2).
\label{eq:dij_linear_U}
\eea
Therefore, the repulsive inter-shell interactions always lower the scaling dimensions of the inter-shell tunneling terms, at least for small $U$.
In particular, this indicates that the inter-shell tunnelings are rendered more relevant at low energies in the RG sense.

\section{Low energy theory with inter-shell tunnelings}
\label{sec:low_energy_field_theory}

\subsection{The gapless modes of center of mass motions}
\label{sec:center_of_mass}

Now we are prepared to discuss the effects of inter-shell tunnelings.
In general, some of the tunneling operators are relevant, while some are irrelevant.
We will build up the low energy theory for $H=H_0+H_{\text{int}}+H_T$ by integrating over the modes which are rendered massive by the relevant tunneling terms.
As a consequence, the number of Luttinger liquid channels at low energies is reduced.

Since $K_i\sim 6$  ($1\leq i\leq P$), \cite{Maestro2011}  the value of $d_{ij}$ in Eq. (\ref{eq:dij_linear_U}) is around $3$ in the absence of inter-shell interactions.
According to Eq. (\ref{eq:dij_linear_U}),  repulsive interactions always lower  the scaling dimensions $d_{ij}$.
If $d_{ij}$ becomes smaller than $2$, then the corresponding tunneling $\cos(\phi_i-\phi_j)$ is relevant and flows to the strong coupling limit at low energies.
Graphically, as shown in Fig. \ref{fig:grouped_channels}, we connect the two channels by a solid line if the tunneling term between them is a relevant operator.
In this way, the $P$ channels can be partitioned into $Q$ ($\leq P$) groups.
Within each group, any two channels are connected by a path formed by the solid lines,
whereas for two channels in two different groups, there is no path connecting them. 

Let's consider the $i$'th group containing $P_i$ channels,
where $\sum_{i=1}^QP_i=P$.
An example is shown in Fig. \ref{fig:grouped_channels}, in which $P_1=5$, $P_2=3$, $P_3=1$, and $P=9$, $Q=3$.
Let $\{i_{1},...,i_{P_i}\}$ be the numberings of the channels in the $i$'th group.
Then in the strong coupling limit, the tunneling potential becomes
\bea
\sum_{1\leq k<l\leq P_i}T_{i_ki_l}\cos(\phi_{i_k}-\phi_{i_l}),
\label{eq:strong_coupling_T}
\eea
in which we have denoted $T_{i_ki_l}=b^{2-d_{i_ki_l}}t_{i_ki_l}$ as the  RG flowed coupling at low energies when the cutoff is reduced by a factor of $b$.
We note that not all $T_{i_ki_l}$'s are nonzero.
If $d_{i_ki_l}$ is larger than $2$, then the corresponding $T_{i_ki_l}$ vanishes.
However, by assumption, any two channels within $\{i_{1},...,i_{P_i}\}$ can be connected by a path of nonzero $T_{i_ki_l}$'s. 
We also note  that $T_{i_ki_l}$ can be either positive or negative depending on the sign of the bare tunneling term $t_{i_ki_l}$.
The strategy is to perform a mean field (i.e., classical) analysis to the RG flowed potential in Eq. (\ref{eq:strong_coupling_T}).
In the strong coupling limit, the ground state of the system is determined by minimizing the potential in Eq. (\ref{eq:strong_coupling_T}).
The simplest situation is when all $T_{i_ki_l}$'s are negative.
Then the minimum solution is given by $\phi_{i_k}\equiv \phi_{i}^{(0)}$ where $1\leq k\leq P_i$ and $\phi_i^{(0)}$ is some arbitrary real number.

For general $T_{i_k,i_l}$'s, we assume that $\phi_{i_k}=\phi_{i_k}^{(0)}$ ($1\leq k \leq P_i$) is a minimum solution.
Apparently, translating all $\phi_{i_l}$'s by the same amount does not cost any energy,
since the cosine potential only depends on the difference $\phi_{i_k}-\phi_{i_l}$.
Therefore, the shifted coordinates
\bea
\phi_{i_k}=\phi_{i_k}^{(0)}+\lambda, ~\lambda\in \mathbb{R}
\label{eq:center_shift}
\eea
 also minimizes the potential.
Hence, the shift of an overall phase is a gapless mode,
and it corresponds to the center of mass motion of all the $P_i$ channels within the $i$'th group. 

Supposing  we have found a minimum solution of the tunneling potential in the strong coupling limit for each group of channels,
next we expand Eq. (\ref{eq:strong_coupling_T}) around the minimum solutions.
Let $\delta \phi_{i_l}$ defined as
\bea
\delta \phi_{i_l}=\phi_{i_l}-\phi_{i_l}^{(0)}
\eea
be the coordinate parametrizing the deviation from the minimum solution.
Then the tunneling potential can be expanded in a Taylor expansion of $\delta \phi_{i_l}$.
The linear terms vanish since $\{\phi_{i_l}^{(0)}\}_{1\leq l\leq P_i}$ constitutes a saddle point.
Keeping only the quadratic terms, the tunneling potential becomes
\bea
\frac{1}{2\pi} \delta \phi^T M \delta \phi,
\label{eq:M_potential}
\eea
in which $M$ is a $P\times P$ symmetric and semi-positive-definite matrix.

Notice that $M$ contains $Q$ zero eigenvalues, corresponding to translating all the $\phi_j$'s within the same group of channels by a same amount of displacement.
More explicitly, the vector $w_k$ defined as
\bea
w_k=(0,...,1,0,...,1,...0)^T
\label{eq:wk}
\eea
is a null vector of $M$ (i.e., $Mw_k=0$),
in which the ``$1$"'s appear at the $k_1,...,k_{P_k}$ positions.
The massive modes in Eq. (\ref{eq:M_potential}) can be integrated out. 
Hence, at low energies, it is enough to keep the $Q$ gapless modes.
Our next step is to write down the low energy theory for these $Q$ Luttinger liquid modes,
which will be discussed in Secs. \ref{sec:zero_q},\ref{sec:nonzero_q}.
We will first diagonalize the Hamiltonian for the $q=0$ sector.
Then  a nonzero wavevector can be included by a  $k\cdot p$ perturbation on the $Q$ gapless modes in the $q=0$ case.

Here we make a comment on the choice of decomposing the channels.
Although we have based our discussions on a shell model of coupled Luttinger liquids,
it can be readily observed that the whole discussion does not rely how the channels are defined.
For example, one can define the channels according to the angular momentum decomposition of the wavefunctions in a cylindrical geometry.
In that case,  the regrouping of channels discussed in this section due to inter-channel tunnelings equally applies.
The subsequent discussions in Secs. \ref{sec:zero_q}, \ref{sec:nonzero_q}, \ref{sec:substrate} essentially  only rely on  a collection of regrouped channels, not dependent on how these regrouped entities arise.
Hence, our analysis is based on a flexible scheme which captures the overall features and is not sensitive to the microscopic details.

\subsection{The zero wavevector Hamiltonian}
\label{sec:zero_q}

We first consider the $q=0$ case.
The Hamiltonian is given by
\bea
H(q=0)=\frac{1}{2\pi} \nabla \theta^T(q=0) (V_N+\frac{1}{\pi}U)\nabla \theta(q=0)\nn\\
+\frac{1}{2\pi}\delta \phi^T(q=0) M \delta\phi(q=0),
\eea
in which $\nabla \theta_j(q=0)$ is the canonical conjugate partner of $\delta\phi_j(q=0)$.
In what follows, we will drop $q=0$ for simplification of notations.
To diagonalize $H$, we first diagonalize $V_N+\frac{1}{\pi}U$, then rescale it to an identity matrix,
and finally diagonalize $M$.

The real symmetric matrix $V_N+\frac{1}{\pi}U$ can be diagonalized by an orthogonal matrix $O_1$ as
\bea
V_N+\frac{1}{\pi}U=O_1 A_\theta O_1^T,
\label{eq:A_theta}
\eea
in which $A_\theta$ is a diagonal matrix.
Define the transformed coordinates $\theta^{(1)}$ and $\delta\phi^{(1)}$ as
$\theta^{(1)} = O_1^T \theta$, $\delta\phi^{(1)} = O_1^T \delta \phi$.
Then the $\theta$-part in Hamiltonian is diagonalized with matrix kernel $A_\theta$.
Next rescale $\theta^{(1)},\delta\phi^{(1)}$ according to
$\theta^{(2)}=A_\theta^{1/2} \theta^{(1)}$, $\delta \phi^{(2)}=A_\theta^{-1/2} \delta \phi^{(1)}$,
then 
$H=\frac{1}{2\pi} \nabla \theta^{(2),T}\nabla \theta^{(2)}
+\frac{1}{2\pi}\delta \phi^{(2)} \tilde{M} \delta\phi^{(2)}$,
where
\bea
\tilde{M}=A_\theta^{1/2} O_1^T M O_1 A_\theta^{1/2}.
\label{eq:tildeM}
\eea
Since $\tilde{M}$ is symmetric, it can be diagonalized by an orthogonal matrix $O_2$, as
\bea
\tilde{M}=O_2\Lambda_\phi O_2^T,
\eea
where $\Lambda_\phi$ is diagonal.
Define
$\tilde{\theta}=O_2^T \theta^{(2)}$, $\delta \tilde{\phi}=O_2^T \delta \phi^{(2)}$,
we obtain
\bea
H=\frac{1}{2\pi} \nabla \tilde{\theta}^{T}\nabla \tilde{\theta}
+\frac{1}{2\pi}  \delta \tilde{\phi}^{T} \Lambda_\phi  \delta \tilde{\phi}.
\label{eq:H_q0}
\eea

In summary, under the transformations
\bea
\tilde{\theta}&=&O_2^T A_\theta^{1/2} O_1^T \theta,\nn\\
\delta \tilde{\phi}&=&O_2^T A_\theta^{-1/2} O_1^T \delta \phi,
\label{eq:final_transformation}
\eea
the Hamiltonian at $q=0$ is transformed into Eq. (\ref{eq:H_q0}).
In what follows, for $\Lambda_\phi$, we will take the convention of arranging the zero eigenvalues in the upper-left block,
and put the remaining massive eigenvalues to the later positions on the diagonal line, i.e.,
\bea
\Lambda_\phi=\left(\begin{array}{cccccc}
0 & & &&&\\
&...&&&&\\
&&0&&&\\
&&&m_{1}&&\\
&&&&...&\\
&&&&&m_{P-Q}
\end{array}\right),
\label{eq:Lambda_mat}
\eea
in which there are $Q$ zeros among the diagonal elements.
We are going to relate the $Q$ canonical pairs of the collective gapless modes $\{\tilde{\theta}_k,\delta\tilde{\phi}_k \}_{1\leq k \leq Q}$  with the coordinates $\{\theta_j$, $\delta\phi_j\}_{1\leq j\leq P}$, which will be used in deriving the $q\neq 0$ Hamiltonian.

Before proceeding on, let's try to gain a better understanding of the structure of $O_2$.
If $\Phi_k$ is a null vector of $M$,  then $\tilde{\Phi}_k$ given by
\bea
\tilde{\Phi}_k=A_\theta^{-1/2}O_1^T\Phi_k
\label{eq:tildePhik}
\eea
must be a null vector of $\tilde{M}$ (as defined in Eq. (\ref{eq:tildeM})).
We emphasize that this is not true for the eigenvectors of other eigenvalues, i.e.,
if $\Phi$ is an eigenvector of $M$ with a nonzero eigenvalue, then $\Lambda_\theta^{-1/2}O_1^T\Phi$ may not necessarily be an eigenvector of $\tilde{M}$.
By assuming $\Phi_k$ to be the ``center of mass" motion of the $k$'th  group of channels as discussed in Eq. (\ref{eq:center_shift}),
it is clear that $\Phi_k\propto w_k$ where $w_k$ is defined in Eq. (\ref{eq:wk}).
To determine the normalization of $\Phi_k$,
notice that $\tilde{\Phi}_k$ is a column of the orthogonal matrix $O_2$, 
hence $\tilde{\Phi}_k$ is normalized to $1$, i.e., $\tilde{\Phi}_k^T\tilde{\Phi}_k=1$.
This fixes the normalization of $\Phi_k$ to be
\bea
\Phi_k=\frac{1}{\sqrt{\sum_{l=1}^{P_k} [(A_{\theta})_{k_lk_l}]^{-1} }}(\hat{e}_{k_1}+...\hat{e}_{k_{P_k}}),
\label{eq:Normalization_Phi_k}
\eea
in which $e_j$ and $A_\theta$ are defined in Eq. (\ref{eq:e_i}) and Eq. (\ref{eq:A_theta}), respectively,
and $(A_{\theta})_{k_lk_l}$ represents the matrix element of $A_{\theta}$ at the $(k_l,k_l)$ position. 
Since $\tilde{M}$ is diagonalized by $O_2$,
we see that the $k$'th column ($1\leq k \leq Q$) of $O_2$ is $\tilde{\Phi}_k$, i.e.,
\bea
(O_2)^{cl}_k=\tilde{\Phi}_k
\label{eq:O2cl}
\eea
in which $(C)^{cl}_k$ denotes the column vector formed by the $k$'th column of the matrix $C$.
More explicitly,
\bea
O_2=(\tilde{\Phi}_1,...\tilde{\Phi}_k, \tilde{\Phi}^\prime_{k+1}...\tilde{\Phi}^\prime_{P}),
\eea
in which $\tilde{\Phi}^\prime_j$ ($k+1\leq j \leq P$)  are the eigenvectors of the massive eigenvalues in Eq. (\ref{eq:Lambda_mat}).

Next, we express $\tilde{\theta}_k$ ($1\leq k \leq Q$) -- which is the gapless mode of the $k$'th component of the column vector $\tilde{\theta}$ -- in terms of $\theta_i$'s ($1 \leq i \leq P$).
According to Eq. (\ref{eq:final_transformation}), $\tilde{\theta}_k$ is equal to
$[(O_2)_{k}^{cl}]^T A_\theta^{1/2}O_1^T \theta$.
Using Eqs. (\ref{eq:tildePhik},\ref{eq:O2cl}), it is straightforward to obtain
\bea
\tilde{\theta}_k=\Phi_k^T \theta.
\label{eq:theta_to_tildetheta}
\eea
By virtue of Eq. (\ref{eq:Normalization_Phi_k}), we conclude that for the ``center of mass" motion of  the $k$'th group of channels $\{j_1,...j_{P_k}\}$,
the gapless mode is given by $\tilde{\theta}_k\propto \theta_{k_1}+...+\theta_{k_{P_k}}$.
Here we make a comment on $(A_\theta)_{k_lk_l}$'s which appear in the normalization factor of $\Phi_k$.
According to Eq. (\ref{eq:A_theta}), $(A_{\theta})_{k_lk_l}$ is equal to $v_{Nk_l}$ up to lowest order in $U$.
Since the diagonal elements of $U$ vanish,
the first order corrections of the eigenvalues of $A_\theta$ are zero.
Hence, the next order term in $(A_{\theta})_{k_lk_l}$ is in the order of $U^2$, i.e.,
\bea
(A_{\theta})_{k_lk_l}=v_{Nk_l}+O(U^2).
\eea

We also examine $\delta\tilde{\phi}$ and derive the component of $\delta \phi_i$ on $\delta\tilde{\phi}_k$ ($1\leq k \leq Q$).
Notice that 
$\delta \phi = O_1A_\theta^{1/2} O_2 \delta \tilde{\phi}$.
Thus the component of $\delta\phi$ on $\delta \tilde{\phi}_k$ is given by the $k$'th column of $O_1A_\theta^{1/2} O_2$.
On the other hand, 
$(O_1A_\theta^{1/2} O_2)^{cl}_k=(O_1A_\theta^{1/2}) (O_2)^{cl}_k=\Phi_k$,
where Eqs. (\ref{eq:tildePhik},\ref{eq:O2cl}) are used.
This shows that the component of $\delta \phi$ on $\delta \tilde{\phi}_k$ is given by $\Phi_k$.
Taken into account the normalization, we obtain
\bea
\delta \phi=\sum_{k=1}^Q\frac{1}{\sqrt{\sum_{l=1}^{P_k} [(A_\theta)_{k_lk_l}]^{-1} }} (\hat{e}_{k_1}+...\hat{e}_{k_{P_k}})\delta \tilde{\phi}_k\nn\\
+ \text{massive modes},
\label{eq:tildephi_to_phi}
\eea
in which the notation ``massive modes"  in Eq. (\ref{eq:tildephi_to_phi}) denote the contributions from the massive eigenvectors $\delta\tilde{\phi}_k$ ($Q+1\leq k\leq P$).

In summary, the transformations between $\{\tilde{\theta},\delta\tilde{\phi}\}$ and $\{\theta,\delta\phi\}$ are given by Eqs. (\ref{eq:theta_to_tildetheta},\ref{eq:tildephi_to_phi}),
and  can be arranged into the following matrix forms
\bea
\tilde{\theta}^L&=&\left(\begin{array}{c}
\Phi_1^T\\
...\\
\Phi_k^T
\end{array}\right)\theta,\nn\\
\delta \phi&=&(\Phi_1,...,\Phi_k)\delta \tilde{\phi}^L+\text{massive modes},
\label{eq:tilde_to_nontilde_mat}
\eea
 in which $\tilde{\theta}^L$ and $\delta \tilde{\phi}^L$ are both $Q$-component column vectors
 defined as 
 \bea
 \tilde{\theta}^L&=&( \tilde{\theta}_1,..., \tilde{\theta}_Q)^T\nn\\
 \delta \tilde{\phi}^L&=& (\delta \tilde{\phi}_1^L,...,\delta \tilde{\phi}_Q^L)^T,
 \eea
 where $\tilde{\theta}^L_k,\delta \tilde{\phi}_k^L$ ($1\leq k\leq Q$) are used to denote the gapless Luttinger liquid modes within $\tilde{\theta}_j,\delta \tilde{\phi}_j$ ($1\leq j\leq P$).
 The components of $\delta \phi$ on $\delta \tilde{\phi}_l$ ($Q+1\leq l\leq P$) are abbreviated in the notation ``massive modes" and not explicitly shown.

Finally we note that besides the detailed derivations of the transformations in Eq. (\ref{eq:tilde_to_nontilde_mat}) given within this section,
there are understandings of Eq. (\ref{eq:tilde_to_nontilde_mat}) based on considerations on general grounds.
An understanding of Eq. (\ref{eq:tilde_to_nontilde_mat}) from the point of view of the Noether theorem is discussed in Appendix \ref{sec:Noether},
which in particular, does not rely on the Gaussian fluctuation approximation made in Eq. (\ref{eq:M_potential}).
In addition, the transformation from $\delta \phi$ to $\delta \tilde{\phi}$ can be inferred from that from $\theta$ to $\tilde{\theta}$ as discussed in Appendix \ref{sec:canonical}, since the two transformations together constitute a canonical transformation.

\subsection{The nonzero wavevector Hamiltonian}
\label{sec:nonzero_q}

Now we are able to write down the low energy theory for the $Q$ gapless modes by including nonzero wavevectors,
which can be achieved using a $k\cdot p$ perturbation theory.
Comparing the Hamiltonians  between the $q=0$ and $q\neq 0$ cases,
we see that there is one additional term for a nonzero $q$ which involves the derivatives of $\delta \phi$, i.e.,
\bea
\Delta H(q)=\frac{1}{2\pi} \nabla \delta \phi^T(q)V_J\nabla\delta \phi(-q),
\label{eq:Ham_phi_nonzero_q}
\eea
in which  $\nabla \delta \phi=\nabla \phi$ is used and $V_J$ is defined in Eq. (\ref{eq:VN_VJ}).
Notice that in the $k\cdot p$ treatment, we should replace $\nabla$ in Eq. (\ref{eq:tilde_to_nontilde_mat}) by $\pm iq$,
but we choose to keep the gradient symbol for simplicity.

By integrating out the massive modes, it is enough to keep the gapless modes $\delta \tilde{\phi}^L_k$ ($1\leq k\leq Q$) in Eq. (\ref{eq:Ham_phi_nonzero_q}).
Using Eq. (\ref{eq:tilde_to_nontilde_mat}), 
we obtain
\bea
\Delta H(q)=\frac{1}{2\pi} \nabla \delta \tilde{\phi}^{L,T}(q)\tilde{V}_J\nabla\delta \tilde{\phi}^L(-q),
\eea
in which
\bea
\tilde{V}_J=
\left(\begin{array}{c}
\Phi_1^T\\
...\\
\Phi_k^T
\end{array}\right)
V_J(\Phi_1,...,\Phi_k).
\label{eq:VJtilde}
\eea
Since $V_J$ is diagonal, and different $\Phi_k$'s do not have any common channel,
it is clear that $\tilde{V}_J$ is diagonal, i.e.,
\bea
\Phi_m^T V_J \Phi_n = (\tilde{V}_J)_{mm}\delta_{mn},~1\leq m,n\leq Q.
\eea
The normalization factor can be straightforwardly calculated as
\bea
(\tilde{V}_J)_{kk}=\frac{\sum_{l=1}^{P_k} v_{Jk_l}}{\sum_{l=1}^{P_k} (A_{\theta})_{k_lk_l}^{-1}},
\label{eq:tilde_VJkk}
\eea
in which $A_\theta$ is defined in Eq. (\ref{eq:A_theta}).
Keeping only the $O(1)$ terms, we have
\bea
(\tilde{V}_J)_{kk}= \frac{\sum_{l=1}^{P_k} v_{Jk_l}}{\sum_{l=1}^{P_k} v_{Nk_l}^{-1}} +O(U^2).
\eea

In summary, the low energy theory for the $Q$ gapless modes is 
\bea
\tilde{H}=\frac{1}{2\pi}\sum_{k=1}^Q\int dx  \big[ \nabla \tilde{\theta}^L_k\nabla \tilde{\theta}^L_k
+\tilde{v}_{Jk} \nabla \delta \tilde{\phi}^{L,T}_k  \nabla \delta \tilde{\phi}^L_k\big],
\label{eq:Q_Luttinger}
\eea
in which the velocity and Luttinger parameter for the $k$'th mode are
\bea
\tilde{v}_{Jk}=(\tilde{V}_J)_{kk}, ~\tilde{K}_k=\frac{1}{\sqrt{\tilde{v}_{Jk}}}.
\label{eq:tildevJk}
\eea

\section{Substrate potential}
\label{sec:substrate}

To understand the fate of the $Q$ remaining gapless boson fields in Eq. (\ref{eq:Q_Luttinger}) which are not pinned by inter-shell tunnelings, we must finally consider the effect of the substrate potential.

The bosonized form of the substrate potential is given in Eq. (\ref{eq:H_S}).
If the substrate potential has a Fourier component at wave-vector $2\pi\rho_{0j}$, 
then the operator $\cos[2\theta_j(x)]$  will appear in the effective Hamiltonian and will be relevant if it has dimension $d_{Sj}<2$. 
Alternatively, if the substrate potential has a random component at this wavevector, $H_S$ will be relevant  if \cite{Giamarchi1988}  $d_{Sj}< 3/2$.
However, to study the RG behaviour
of $H_S$ we must take into account the effects of the inter-shell tunnelings. 
These lead to competing phases since $H_S$
attempts to pin the $\theta_j$ variables whereas $H_T$ attempts to pin the $\phi_j$  variables. 
When a field is pinned,
its dual field fluctuates strongly making the corresponding interaction irrelevant.
Here we assume that the substrate potential is sufficiently weak compared to the inter-shell tunneling, such that Eq. (\ref{eq:H_S}) can be treated as a perturbation on Eq. (\ref{eq:tildevJk}).
This assumption should be true at least for the channels in the central region of the nanopore when the radius of the pore is large.

Rewriting $\cos(2\theta_j)$ in the $\tilde{\theta}$-basis always involves some of the $\tilde{\theta}_k$ variables with $k>Q$ (corresponding to massive modes) which makes $H_T$ irrelevant. 
Thus inter-shell tunneling can stabilize the system against disorder. 
However, we must consider higher order processes which can be relevant. 
For example, the following term which involves the $k$'th gapless collective mode ($1\leq k\leq Q$) in Eq. (\ref{eq:Q_Luttinger}) is allowed via a $P_k$'th order perturbation,
\bea
H_S^{(i)}= \int dx V(x)  \cos[2\pi \rho_{0,(k)}x-2\sum_{l=1}^{P_k} \theta_{k_l} ],
\eea
in which $\rho_{0,(k)}=\sum_{l=1}^{P_k}\rho_{0k_l}$ is the total linear density in the $k$'th group of channels.
Taking into account the fact that $V(x)$ may have a Fourier mode at $2\pi \rho_{0,(k)}$ and a random component, 
the relevance of $H_S^{(k)}$ is determined by the scaling dimension $d_{S,(k)}$ of $\cos[2\sum_{l=1}^{P_k} \theta_{k_l} ]$.

Using Eq. (\ref{eq:tilde_to_nontilde_mat}), $\cos[2\sum_{l=1}^{P_k} \theta_{k_l} ]$ can be rewritten as 
\bea
\cos[2\sum_{l=1}^{P_k} \theta_{k_l} ]=\cos(2\mu_k\tilde{\theta}_k),
\eea
in which $\mu_k$ is given by
$\mu_k=\sqrt{\sum_{l=1}^{P_k} [(A_{\theta})_{k_lk_l}]^{-1}}$.
Therefore, the scaling dimension $d_{S,(k)}$ can be determined as
$d_{S,k}=\mu_k/\tilde{K}_k$.
According to Eq. (\ref{eq:tildevJk}), we obtain
\bea
d_{S,k}=\sqrt{ \big(\sum_{l=1}^{P_k} v_{Jj_l}\big)\big(\sum_{l=1}^{P_k} [(A_{\theta})_{k_lk_l}]^{-1}\big) }.
\label{eq:tilde_d_k}
\eea

To understand Eq. (\ref{eq:tilde_d_k}),
let's consider the special case of identical $P$ channels of Luttinger liquids without inter-shell interactions, i.e.,
\bea
v_{Ji}\equiv v_J, ~v_{Ni}\equiv v_N,
\label{eq:identical_channels}
\eea
where $1\leq i\leq P$.
Then it is clear that up to $O(U^0)$, we have
\bea
d_{S,k}=\sqrt{P_k^2 \frac{v_J}{v_N}}=\frac{P_k}{K},
\label{eq:tildedk_simplified}
\eea
in which $K=\sqrt{\frac{v_J}{v_N}}\sim 6$ is the Luttinger parameter of a single channel in the initial model of $P$ coupled Luttinger liquid channels.
Thus we see that the larger $P_k$ is, the more robust the $k$'th gapless mode in Eq. (\ref{eq:Q_Luttinger}) becomes with respect to disorder effects.

Finally, let's consider an example for illustration, which might be relevant to real situations with a large pore radius.
For simplification, suppose initially there are $P$ approximately identical channels satisfying Eq. (\ref{eq:identical_channels}).
Assume that the inner $P-1$ channels are bound by inter-shell tunnelings and the outer $P$'th channel is left decoupled. 
In this case, we have $Q=2$.
Then according to Eq. (\ref{eq:tildedk_simplified}), the scaling dimensions of the disorder potentials are given by 
\bea
d_{S,1}\sim\frac{P-1}{K},~d_{S,2}\sim\frac{1}{K},
\eea
where $K\sim 6$.
Clearly, the outermost $P$'th shell is localized by the disorder which coats the pore wall.
For the inner $P-1$ shells, they are localized by the disordered substrate potential when $P$ is small.
However, $d_{S,1}$ can be made arbitrarily large by increasing $P$,
hence the effect of disorder potential on the inner entity of the $P-1$ shells
will be made irrelevant for sufficiently large $P$.
As a result, we should be able to observe a Luttinger liquid flowing through the nanopore.

We note that the above analysis provides an understanding to the physical arguments about the 1D to 3D crossover behavior as discussed in Sec. \ref{sec:intro}.
The $P$'th shell is pinned by the wall potential and shields the inner fluids from the substrate such that superfluidity  (here quasi-long ranged superfluidity) is maintained in the central regions.
For more complicated situations, we expect that as long as the pore radius is large enough, there exists a group of shells which are bound together by inter-shell tunnelings, making them robust to disorder effects.
Therefore, a Luttinger liquid channel always exists in the system for nanopores with a large enough radius.

\section{Conclusion}
\label{sec:conclusion}

In conclusion, based on earlier QMC observations, we propose a shell model of coupled Luttinger liquids to describe the helium-4 mass flow through rough-walled nanopores.
Using this shell model, the effects of substrate potential and increasing pore radius are studied.
For small pore radius, all helium-4 atoms are localized by the substrate potential. 
However, at a critical radius, a single component gapless Luttinger liquid emerges as the first step in the crossover to 3D behavior. 
This result is related to the standard picture for larger pores where a layer of bosons near the pore wall smooth out the substrate potential and allow a tube of atoms to flow through the center with zero viscosity.
It suggests that there may be a range of pore radii over which single component Luttinger liquid behavior could be observed. Surprisingly, this does not require such a small radius that there is only one shell. 
Rather the minimum required pore radius corresponds to multiple shells in order for the effects of the substrate potential to be screened.
Numerical  test of the proposed scenario is worth further studies.\\

{\it Acknowledgments}
We thank A. Del Maestro and S. Sahoo for helpful discussions.
WY and IA acknowledge support from NSERC Discovery Grant 04033-2016.

\appendix

\section{Proof of Eq. (\ref{eq:D_integral})}
\label{sec:Proof_lemma}

In this appendix, following Ref. \onlinecite{Magnus1988}, we give a quick proof of Eq. (\ref{eq:D_integral}).
Let $A(\lambda)=A+\lambda B$. 
Let $D$ be defined as
\bea
D=[\frac{d}{d\lambda} \sqrt{A+\lambda B}]_{\lambda=0},
\eea
i.e., $D=[\frac{d}{d\lambda} \sqrt{A(\lambda)}]_{\lambda=0}$.
Differentiating $[\sqrt{A(\lambda)}]^2=A(\lambda)$, we have 
\bea
D\sqrt{A}+\sqrt{A} D =[\frac{d}{d\lambda}A(\lambda)]_{\lambda=0}=B,
\label{eq:differential_D}
\eea
which has a unique solution for $D$.
We show that the integral expression for $D$ in Eq. (\ref{eq:D_integral}) satisfies Eq. (\ref{eq:differential_D}).
In fact,
\begin{flalign}
&D\sqrt{A}+\sqrt{A}D  \nn\\
&= \int_0^\infty dt \big[
e^{-t\sqrt{A}}Be^{-t\sqrt{A}}\sqrt{A}+\sqrt{A}e^{-t\sqrt{A}}Be^{-t\sqrt{A}}
\big]\nn\\
&=\int_0^\infty dt(-\frac{d}{dt})(e^{-t\sqrt{A}}Be^{-t\sqrt{A}})\nn\\
&= B,
\end{flalign}
completing the proof of Eq. (\ref{eq:D_integral}).

\section{Noether theorem and the collective gapless modes}
\label{sec:Noether}

For simplification, we consider the special case of $Q=1$, i.e.,
there is only one gapless mode, which corresponds to the center of mass motion of all the $P$ channels.
The general case of an arbitrary $Q$ can be discussed in a similar manner by considering the channels of each collective gapless mode separately.  

To apply the Noether theorem, we consider the original Hamiltonian in Eq. (\ref{eq:H}) in its bosonized form.
The system has a continuous symmetry defined as 
\bea
\phi_i\rightarrow \phi_i+\lambda,
\eea
where $1\leq i\leq P$, and $\lambda\in \mathbb{R}$.
The corresponding Noether current is 
\bea
j^\mu=\sum_i\frac{\partial \mathcal{L}}{\partial(\partial_\mu \phi_i)},
\eea
in which $\mu=0$ and $1$ corresponding to the time and spatial coordinates, respectively, and  the Lagrangian $\mathcal{L}$ density is given by
\bea
&\mathcal{L}=\frac{1}{\pi}\sum_i \partial_t\phi_i \partial_x \theta_i-
\frac{1}{2\pi}  \sum_i [ v_{Ji} (\partial_x\phi_i)^2+v_{Ni} (\partial_x\theta_i)^2]\nn\\
&-\frac{1}{\pi^2} \sum_{i<j}  \hat{U}_{ij}\partial_x\theta_i\partial_x\theta_j-\sum_{i,j}t_{ij}\cos(\phi_i-\phi_j).\nn\\
\eea
This gives
\bea
j^0&=&\frac{1}{\pi}\sum_i \partial_x\theta_i,\nn\\
j^1&=&-\frac{1}{\pi}\sum_i v_{Ji}\partial_x \phi_i.
\eea
The local conservation law 
\bea
\partial_\mu j^\mu=0
\eea
then implies 
\bea
\partial_t\sum_i \partial_x\theta_i=\sum_i v_{Ji}\partial_x^2 \phi_i.
\label{eq:Noether_eq}
\eea

Multiplying both sides of Eq. (\ref{eq:Noether_eq}) by $\Phi_1^T$, where, according to Eq. (\ref{eq:Normalization_Phi_k}), $\Phi_1$ is defined as 
\bea
\Phi_1=\frac{1}{\sqrt{\sum_{l=1}^{P} [(A_{\theta})_{ll}]^{-1} }} (1,...,1)^T,
\eea
Eq. (\ref{eq:Noether_eq}) can then be alternatively written as
\bea
\partial_t\partial_x\tilde{\theta}_1=\Phi^TV_J \partial_x^2 \phi,
\label{eq:Nother_eq_2}
\eea
in which $V_J$ is the matrix defined in Eq. (\ref{eq:VN_VJ}), and $\tilde{\theta}_1$ represents the collective coordinate of the center of mass motion defined in Eq. (\ref{eq:tilde_to_nontilde_mat}).
Notice that the designation of the coordinate as $\tilde{\theta}_1$ is consistent with the convention taken in Eq. (\ref{eq:Lambda_mat}),
where the numbering of the gapless modes are in front of the massive modes.
Using  Eq. (\ref{eq:tilde_to_nontilde_mat}),  $\partial_x^2 \phi$ can be expressed in terms of $\partial_x^2\delta \tilde{\phi}_i$'s.
Then Eq. (\ref{eq:Nother_eq_2}) becomes
\bea
\partial_t\partial_x\tilde{\theta}_1=\Phi^TV_J\Phi \cdot\partial_x^2 \delta\tilde{\phi}_1+\partial_x^2(\text{massive modes}).
\label{eq:Nother_eq_3}
\eea
According to Eq. (\ref{eq:VJtilde}), Eq. (\ref{eq:Nother_eq_3}) is simply
\bea
\partial_t\partial_x\tilde{\theta}_1=(\tilde{v}_J)_{1}\partial_x^2 \delta\tilde{\phi}_1+\partial_x^2(\text{massive modes}).
\label{eq:Nother_eq_4}
\eea
At low energies, the massive modes can be removed from Eq. (\ref{eq:Nother_eq_3}).
Thus Eq. (\ref{eq:Nother_eq_4}) coincides exactly with the equation of motion for $\partial_x\tilde{\theta}_1$ which can be readily derived from the Hamiltonian  in Eq. (\ref{eq:Q_Luttinger}).
This provides an understanding of Eq. (\ref{eq:Q_Luttinger}) in terms of conservation law and  Noether theorem.

\section{Canonical transformation from $\{\theta,\delta \phi\}$ to $\{\tilde{\theta},\delta \tilde{\phi}\}$}
\label{sec:canonical}

Consider the following linear transformations
\bea
\tilde{\theta}=U_\theta \theta,~\delta\tilde{\phi}=U_\phi \delta \phi,
\label{eq:canonical_app}
\eea
in which both $U_\alpha$ ($\alpha=\theta,\phi$) are $P\times P$ matrices.
Assuming $\{\tilde{\theta},\delta\tilde{\phi}\}$  to satisfy the same commutation relations as $\{\theta,\delta\phi\}$, i.e., Eq. (\ref{eq:commutation}), it is straightforward to obtain
\bea
U_\theta U_\phi^T=I_P,
\eea
where $I_P$ represents the $P\times P$ identity matrix.
Therefore, we obtain $\delta\tilde{\phi}=U_\theta^{-1,T}\delta \phi$, or alternatively,
\bea
\delta \phi=U_\theta^T\delta\tilde{\phi}.
\eea
This is exactly Eq. (\ref{eq:tilde_to_nontilde_mat}).

In summary, according to Appendices \ref{sec:Noether}, 
we see that the expression of $\tilde{\theta}$ is fixed by Noether theorem,
since its spatial derivative $\partial_x\tilde{\theta}$ simply corresponds to the Noether charge.
Then the transformation for $\delta\tilde{\phi}$ is determined from the property of the canonical transformation as discussed in this appendix.  
However, we emphasize that the usefulness of Eq. (\ref{eq:canonical_app}) in diagonalizing the Hamiltonian is based on the Gaussian fluctuation approximation made in Eq. (\ref{eq:M_potential}).


\end{document}